\documentstyle[prl,aps,psfig]{revtex}

\begin{document}

\twocolumn[\hsize\textwidth\columnwidth\hsize\csname
@twocolumnfalse\endcsname

\widetext
\title{Spontaneous plaquette dimerization in the $J_{1}{-}J_{2}$ Heisenberg model}
\author{Luca Capriotti and Sandro Sorella}

\address{ 
Istituto Nazionale di Fisica della Materia and International School for 
Advanced Studies,\\
 Via Beirut 4, I-34013 Trieste, Italy \\ 
} 

\date{\today}
\maketitle

\begin{abstract}
We investigate the non magnetic phase of the 
spin-half frustrated Heisenberg antiferromagnet on the square lattice 
using exact diagonalization (up to 36 sites) and 
quantum Monte Carlo techniques (up to 144 sites). 
The spin gap and the susceptibilities for the most important 
crystal symmetry breaking operators are computed. 
A genuine and somehow unexpected ``plaquette RVB'', with 
spontaneously broken translation symmetry and no broken rotation symmetry, 
comes out from our numerical simulations as the most 
plausible ground state  for $J_2/J_1 \simeq 0.5 $.
\end{abstract}
]

\narrowtext

The nature of the non magnetic phases of a quantum antiferromagnet
is a topic of great interest and has been a subject of intense 
theoretical investigation since Anderson's suggestion 
\cite{anderson} about the possible
connections with the mechanism of high-$T_{c}$ superconductivity. 

Within the Heisenberg model the simplest way in which
the antiferromagnetism can be destabilized is by introducing a
next-nearest-neighbor frustrating interaction leading
to the so called $J_{1}{-}J_{2}$ Hamiltonian
\begin{equation}
\hat{H}=J{_1}\sum_{n.n.}
\hat{{\bf {S}}}_{i} \cdot \hat{{\bf {S}}}_{j}
+ J{_2}\sum_{n.n.n.}
\hat{{\bf {S}}}_{i} \cdot \hat{{\bf {S}}}_{j}~~,
\label{j1j2ham}
\end{equation}
where ${\bf \hat{S}}_{i}=(\hat{S}^x_i,\hat{S}^y_i,\hat{S}^z_i)$ 
are $s-1/2$ operators on a square lattice.
$J_{1}$ and $J_{2}$ are  the (positive) antiferromagnetic superexchange
couplings between nearest and next-nearest-neighbor pairs of spins
respectively.
In the following we will consider finite clusters of $N$ sites with
periodic boundary conditions (tilted by $45^{\circ}$ only for $N=32$) .

Although there is a general consensus about the disappearance
of the N\'eel order in the ground state (GS) of the present model
for $0.38\lesssim J_{2}/J_{1}\lesssim 0.60$ \cite{chandra,schultz,lettersr}, 
no definite conclusion has been drawn on the nature of the non magnetic 
phase yet.
In particular an open question is whether the GS of the
$J_{1}{-}J_{2}$ Heisenberg model is 
a resonating valence bond (RVB) spin liquid
with no broken symmetries, as it was originally 
suggested by Figueirido {\em et al.}
\cite{liquid}. 
The other possibility is a GS which is still ${\cal SU}(2)$ invariant, 
but nonetheless breaks some crystal symmetries, 
dimerizing in some special pattern  \cite{dagotto,gelfand,singh,zith,kotov,columnar}.

In this paper we address this point using 
exact diagonalization (ED) and a quantum Monte Carlo technique, 
the Green function Monte Carlo (GFMC), 
which allows the calculation of GS expectation values on fairly 
large system sizes ($L\le 144$).
This is extremely important to draw reasonable conclusions on the 
physical thermodynamic, zero temperature, properties of the model.

For frustrated spin systems as well as for fermionic models,
quantum Monte Carlo methods are affected by the so called 
{\em sign problem} that can be controlled, at present,
only at the price of introducing some kind of approximations, 
such as the fixed node (FN) one \cite{fn}.
In this work we have also extensively used a recently 
developed technique, the Green
function Monte Carlo with Stochastic Reconfiguration (GFMCSR), 
which improves systematically the accuracy of the FN 
approximation  for GS  calculations \cite{lettersr,gfmcsr,triang,tj}.  

The FN method allows to work without any sign problem by 
using the following simple strategy: 
the exact imaginary time propagator $e^{-\tau \hat{H}}$ -- used 
to filter out the GS from the best variational guess $|\psi_G\rangle$ -- 
is replaced by an approximate propagator $e^{-\tau \hat{H}_{\rm FN}}$ 
such that the nodes of the propagated 
state $e^{-\tau \hat{H}_{\rm FN} } |\psi_G\rangle  $ do not change, 
due to  an  appropriate 
choice of the effective FN Hamiltonian $\hat{H}_{\rm FN}$ 
(which in turn depends on $|\psi_G\rangle$).
The FN approximation becomes exact if the so called {\em guiding} wavefunction 
$|\psi_G\rangle$ is the exact GS.
However for frustrated spin models even the best 
variational wavefunction of the Jastrow type \cite{lettersr}, 
used to guide the FN dynamic, provides rather poor 
results even for the GS energy expectation value \cite{lettersr,gfmcsr,triang}.

The GFMCSR method allows to release the FN approximation and to obtain 
results much less depending on the quality of the  guiding wavefunction. 
During each short imaginary time evolution  $\tau \to  \tau + \Delta \tau$,  
where both the exact and the approximate propagation can be performed without sign problem
instabilities, the FN dynamic is systematically improved by requiring 
that a  given  number $p$ of {\em mixed averages} \cite{gfmcsr} of correlation 
functions are  propagated consistently  with the exact dynamic. 
By increasing the number of correlation functions one typically 
improves the accuracy of the calculation  since the method becomes  exact  
if all the independent correlation functions are included in the 
stochastic reconfiguration (SR) scheme. 

Typically \cite{lettersr,triang}, 
few correlation functions ($p \sim 10$) allow to obtain 
rather accurate values of the GS energy
with an error much less than 1$\%$ -- for lattice sizes ($N\leq 36$) 
where the exact solution is available numerically -- 
and without a sizable loss of accuracy with increasing size.
Such accuracy is usually enough to reproduce some physical features that 
are not contained at the variational level,
as it has been shown in a previous study of the present model \cite{lettersr}.
In the latter work, in fact,  with a gapless guiding wavefunction, 
it has been possible to detect a 
finite spin gap in the thermodynamic limit for $J_{2}/J_{1}\gtrsim 0.4$.

\begin{figure}
\centerline{\psfig{bbllx=70pt,bblly=250pt,bburx=500pt,bbury=550pt,%
figure=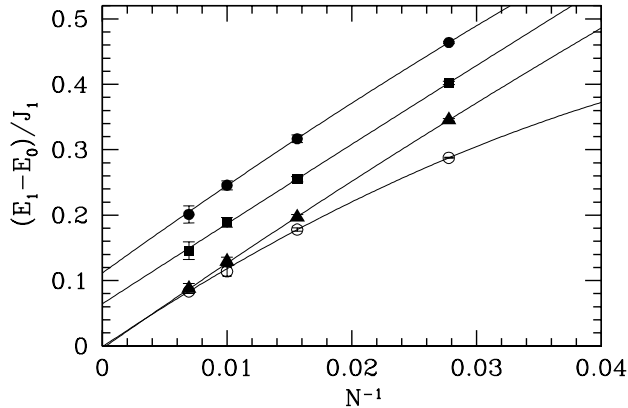,width=80mm,angle=0}}
\caption{\baselineskip .185in \label{gap}
Size scaling of the energy gap to the first $S=1$ spin excitation
obtained with the GFMCSR technique for $J_{2}/J_{1}=0.38$ (full triangles),
0.45 (full squares) and 0.50 (full circles). Data for the unfrustrated ($J_{2}=0
$)
Heisenberg model taken from Ref.~\protect\cite{heis}, are also shown for 
comparison
(empty circles). Lines are weighted quadratic fits of the data.
}
\end{figure}

We have extended the previous GFMCSR calculation, with
the same guiding wavefunction of Ref.~\cite{lettersr}, by including in 
SR conditions not only the energy and
all $\hat{S}^{z}_{i} \hat{S}^{z}_{j}$ independent by symmetry, 
but also the antiferromagnetic order 
parameter. The latter, as discussed in Ref.~\cite{gfmcsr}, 
though not improving the accuracy of the
calculation, allows a very stable and reliable simulation 
for large $p$.
The new results, extended up to $N=144$, 
confirm the previous findings of a finite spin gap 
for $J_{2}/J_{1} \gtrsim 0.40$ (Fig.~\ref{gap}).

As suggested in Refs.\cite{singh,zith,columnar,ST}, in order to investigate 
the possible occurrence of a spontaneously dimerized GS displaying
some kind of crystalline order, we have calculated the response of the system
to operators breaking the most important lattice symmetries. 
This can be done by adding to the Hamiltonian (\ref{j1j2ham})
a term $\delta \hat{O}$, where $\hat{O}$ is an 
operator that breaks some symmetry of $\hat{H}$. 
On a finite size, the GS expectation value of $\hat{O}$
vanishes by symmetry for $\delta=0$ and the GS energy per site 
has corrections proportional to $\delta^2$ 
as by the Hellmann-Feynman theorem
$-de(\delta)/d\delta=\langle \hat{O} \rangle_{\delta}/N$. 
Therefore $e(\delta)\simeq e_0-\chi \delta^2 /2$ , being 
$\chi$ the generalized susceptibility associated to the operator
$\hat{O}$, namely, 
$\chi = 2 \langle \psi_{0}| \hat{O} (E_{0}-\hat{H})^{-1}
\hat{O} | \psi_{0} \rangle/NJ_{1}$.
For $N\to \infty$, if true long range order (LRO) exists 
in the thermodynamic GS,
an infinitesimal field $\delta \sim 1/N$ must give a finite 
$\langle \hat{O} \rangle_{\delta}/N \sim \chi \delta$
implying that the finite size susceptibility 
$\chi = \langle \hat{O} \rangle_{\delta}/\delta N$ has to diverge 
with the system size \cite{lieb}. 
Thus susceptibilities are a very sensitive
tool for detecting the occurrence of LRO. 

We have considered the response of the system
to the following symmetry breaking operators 
\begin{eqnarray}
\hat{O}_{\rm C} &=& \sum_{i} \big( \hat{{\bf {S}}}_{i} \cdot \hat{{\bf {S}}}_{i+x} -
\hat{{\bf {S}}}_{i} \cdot \hat{{\bf {S}}}_{i+y}\big)~, 
\label{column} \\
\hat{O}_{P}&=&\sum_{i} e^{i{\bf Q_0}\cdot{\bf r}_i}
\hat{{\bf {S}}}_{i} \cdot \hat{{\bf {S}}}_{i+x}~, \label{plaq}
\end{eqnarray}
with $x=(1,0)$, $y=(0,1)$ and ${\bf Q_0}=(\pi,0)$,
for the rotation and the translation symmetry, respectively.
Within ED and GFMC technique the susceptibility 
$\chi=-d^{2}e(\delta)/d\,\delta^{2} |_{\delta=0}$ can be evaluated by
computing the GS energy per site in presence of the perturbation 
for few values of $\delta$, 
and by estimating numerically 
the limit $\delta \to 0$ of the quantity 
$\chi(\delta)= -2(e(\delta)-e_0)/\delta^2$.

\begin{figure}
\centerline{\psfig{bbllx=60pt,bblly=235pt,bburx=540pt,bbury=550pt,%
figure=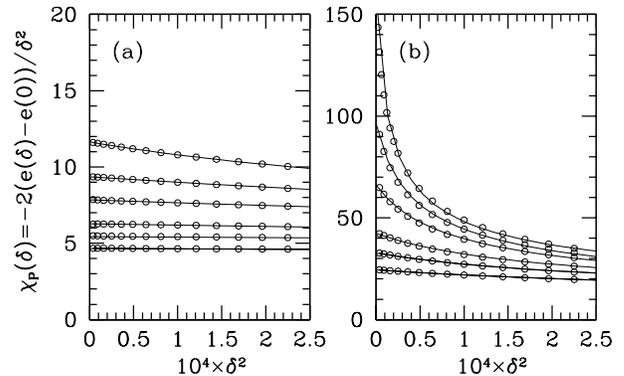,width=80mm,angle=0}}
\caption{\baselineskip .185in \label{chi1d}
ED results for the $J_{1}{-}J_{2}$ chain: 
$\chi_{P}(\delta)$ associated to the operator $\hat{O}_{P}$ 
(breaking the translational invariance) 
for $J_2/J_1=0.2$ (a) and $J_2/J_1=0.4$ (b).
Data are shown for $N=12,14,16,20,24,30$
for increasing values of $\chi_{P}(\delta)$.
}
\end{figure}

As we have tested in the one dimensional $J_1-J_2$ model, 
the numerical study of  LRO by means of $\chi(\delta)$ 
is very effective and reliable. 
Here a quantum critical point at $J_{2}/J_{1}\simeq 0.2412$ 
separating a gapless spin fluid phase from a gapped dimerized GS, 
which is two-fold degenerate, is rather well accepted \cite{1dcrit1,1dcrit2,1dcrit3}.
As shown in Fig.~\ref{chi1d}, the response 
of the system to the perturbation $\delta \hat{O}_{P}$ [Eq.~(\ref{plaq})],
breaking the translation invariance with momentum $k=\pi$,
is very different below and above the dimer-fluid transition point.
However it is extremely important to perform very accurate calculations at 
small $\delta$ to detect the divergence of the susceptibilities for 
large system sizes. 

\begin{figure}
\centerline{\psfig{bbllx=70pt,bblly=250pt,bburx=500pt,bbury=550pt,%
figure=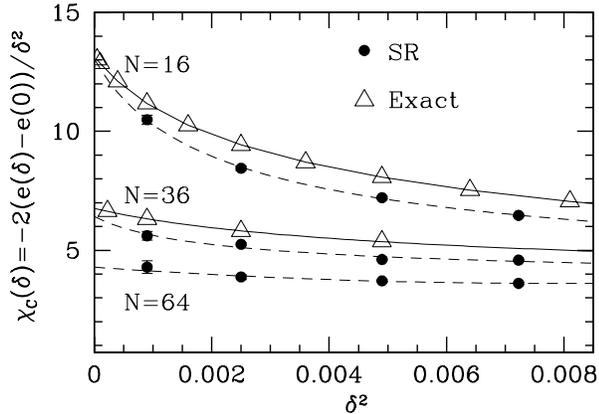,width=80mm,angle=0}}
\caption{\baselineskip .185in \label{chirot}
Exact and GFMCSR calculation of $\chi_C(\delta)$ associated
to $\hat{O}_{C}$ (columnar dimerization)
for $J_2/J_1=0.5$.
}
\end{figure}

In two dimensions,
among the dimerized phases proposed in the literature,
the so-called {\em columnar} and {\em plaquette} RVB 
\cite{dagotto,gelfand,singh,zith,kotov,columnar} are 
the states which have obtained the most convincing 
numerical evidences.
Both the columnar and plaquette RVB break the
translation invariance but only the latter preserves 
the rotation symmetry.
As also suggested in a recent paper by Singh {\em et al.} \cite{columnar},
the appearance of a columnar state can be tested
by using as order parameter 
the operator $\hat{O}_{C}$ defined in Eq.~(\ref{column}). As shown in Fig.~\ref{chirot},
the ED results for $N=16$ and $N=36$ indicate that 
the susceptibility associated with this kind of 
symmetry breaking, $\chi_{C}$, decreases with the system size.
Using the GFMCSR, described before, 
we have extended the calculation up to $N=64$. 
The GFMCSR calculations, which reproduce pretty well
the ED data, rule out clearly the columnar dimer order.

The above result is in disagreement with the conclusions of
several series expansion studies \cite{gelfand,kotov,columnar}.
However, as stated in Ref.~\cite{columnar}, 
the series for $\chi_{C}$ are very irregular and do not allow
a meaningful extrapolation to the exact result.
In our calculation instead, even the ED results for $N<36$,
are already conclusive. 

\begin{figure}
\centerline{\psfig{bbllx=33pt,bblly=250pt,bburx=503pt,bbury=550pt,%
figure=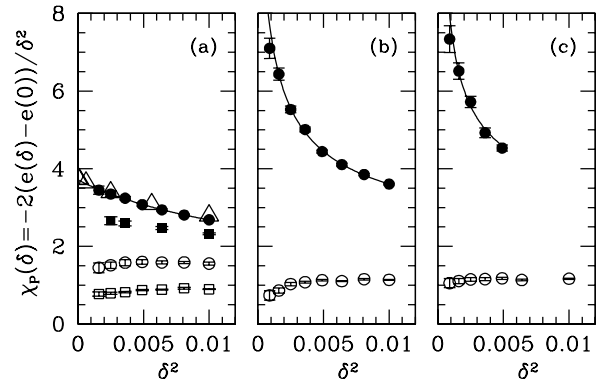,width=80mm,angle=0}}
\caption{\baselineskip .185in \label{comp}
$\chi_{P}(\delta)$ associated to $\hat{O}_{P}$ (plaquette dimerization)
for $J_2/J_1=0.5$, $N=32$ (a), $N=64$ (b) and $N=100$ (c): 
FN (empty squares), GFMCSR (full squares), FN with LS (empty circles), 
GFMCSR with LS (full circles), exact (empty triangles).   
}
\end{figure}

Having established that the columnar susceptibility is bounded,
it is now important to study the response of the $J_{1}-J_{2}$ model 
to a small field coupled to the perturbation $\hat{O}_{P}$ of 
Eq.~(\ref{plaq}), breaking the translation invariance of the 
Hamiltonian.
The evaluation of $\chi_{P}$, with a reasonable accuracy,  
is a much more difficult task.
In fact in this case the ED values of the susceptibility 
for $N=16$ and $N=32$ increase with the size and much
more effort is then required to distinguish if this behavior 
corresponds to a spontaneous symmetry breaking in the thermodynamic limit. 
As it is shown in Fig.~\ref{comp}(a), the FN technique, starting from 
a guiding wavefunction without dimer order, is not able to reproduce
the actual response of the system to $\hat{O}_{P}$,
even on small sizes.
The GFMCSR technique allows to get an estimate of the susceptibility 
which is a factor of three  more accurate, but not  satisfactory enough.
In order to improve this estimate, 
we have attempted to include in the SR conditions many other, 
reasonably simple, correlation functions (such as the spin-spin correlation
 functions
$\hat{{\bf {S}}}_{i} \cdot \hat{{\bf {S}}}_{j}$
for $|r_i-r_j|>\sqrt{2}$), but without obtaining a sizable 
change  of the estimate of $\chi_{P}$. 
In fact the most effective SR conditions are those obtained with 
operators more directly related to the Hamiltonian\cite{gfmcsr,triang}. 

After many unsuccessful attempts, we have realized that 
it is much simpler and straightforward to improve the accuracy 
of the guiding wavefunction itself.  In fact  
it  is reasonable to expect that both the FN and the GFMCSR  will
perform more efficiently with a better $|\psi_G\rangle$, i.e., 
with an improved  initial guess of the GS wavefunction.
This can be obtained by applying a generalized 
Lanczos operator $(1+\alpha\hat{H})$ to the 
variational wavefunction $|\psi_{G}\rangle$, 
where $\alpha$ is a variational parameter. 
This defines the so called {\em one Lanczos step} (LS) wavefunction, 
which has been particularly successful for the $t{-}J$ model
\cite{rice}.

\begin{figure}
\centerline{\psfig{bbllx=70pt,bblly=250pt,bburx=500pt,bbury=630pt,%
figure=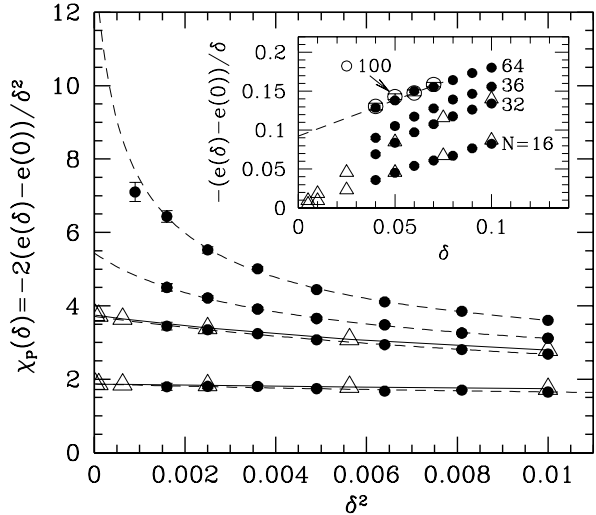,width=80mm,angle=0}}
\caption{\baselineskip .185in \label{chitra}
Exact (empty triangles) and GFMCSR (circles)
calculation of $\chi_P(\delta)$ (plaquette dimerization)
for $J_2/J_1=0.5$ and (from the bottom)
$N=16,32,36,64$. Inset: numerical determination of the order 
parameter (see text). Lines are guides for the eye.
}
\end{figure}

In the present model by using the LS wavefunction,
a clear improvement on the variational estimate of the GS energy is obtained.
More importantly, as shown in Fig.~\ref{comp}(a), the LS wavefunction allows 
a much better estimate of the susceptibility. Remarkably, on all the finite 
sizes  where ED is possible, the GFMCSR estimate of this 
important quantity is basically  exact within few error bars
(see also Fig.~\ref{chitra}).
This calculation 
was  obtained by including in the SR conditions the energy, 
the spin spin correlation functions up to next-nearest-neighbors,
distinguishing also $\hat{S}^{z}_{i} \hat{S}^{z}_{j}$ and 
$(\hat{S}^{x}_{i} \hat{S}^{x}_{j}+\hat{S}^{y}_{i} \hat{S}^{y}_{j})$
($p=4$). 
The mixed averages of these correlation functions can be computed over 
both the wavefunction $|\psi_{G}\rangle$ and the LS wavefunction 
$(1+\alpha\hat{H})|\psi_{G}\rangle$ during the same 
Monte Carlo simulation.
Thus with a LS wavefunction one can also easily double the number of constraints
that are effective to improve the accuracy of the method ($p=8$). 
In this case we have tested that it is irrelevant to add further
long range correlation functions in the SR conditions even for large size.

By increasing the size, the response of the system is very strongly enhanced,
in very close analogy to the one dimensional model 
in the dimerized phase  (see Fig.~\ref{chi1d}(b)).
This is obtained only with the GFMCSR technique, since as shown 
in Fig.~\ref{comp},
 the combination of FN and Lanczos step alone, is not capable to 
detect these strongly enhanced correlations.
For $N=100$ the GFMCSR increases by more than one order of magnitude 
the response of the system to the dimerizing field. 
This effect is particularly striking, considering that the starting guiding  
wavefunction  is spin wave like\cite{franjic}, i.e., gapless, N\'eel ordered
and without any dimer LRO. This suggests that all our systematic 
approximations are able to remove almost completely 
even a very strong bias present at the variational level.

We believe that the numerical results we 
have presented here give  a very robust indication of a spontaneous 
dimerization  with broken translation symmetry but 
without broken  rotation symmetry (as discussed before), i.e., a plaquette RVB. 
This kind of state can be thought of a collection of rotation invariant 
valence bond states 
$$
{\Big |}\hspace{-5pt}\begin{array}{cc}
\begin{picture}(10,30)(-5,-5)
\put(0,0) {\circle{3}}
\put(0,1.8){\line(0,1){15}}
\put(1.9,0){\line(1,0){15}}
\put(0,18){\circle{3}}
\put(1.9,18){\line(1,0){15}}
\end{picture}
&
\begin{picture}(10,30)(-5,-5)
\put(2.1,0) {\circle{3}}
\put(2.1,1.8){\line(0,1){15}}
\put(2.1,18){\circle{3}}
\end{picture}
\end{array} \hspace{-2pt}{\Big \rangle}
= {\Big |}\hspace{-5pt} \begin{array}{cc}
\begin{picture}(10,30)(-5,-5)
\put(0,0) {\circle{3}}
\put(1.9,0){\line(1,0){15}}
\put(0,18){\circle{3}}
\put(1.9,18){\line(1,0){15}}
\end{picture}
&
\begin{picture}(10,30)(-5,-5)
\put(2.1,0) {\circle{3}}
\put(2.1,18){\circle{3}}
\end{picture}
\end{array}\hspace{-2pt}{\Big \rangle}
+
{\Big |}\hspace{-5pt} \begin{array}{cc} 
\begin{picture}(10,30)(-5,-5) 
\put(0,0) {\circle{3}}
\put(0,1.8){\line(0,1){15}}
\put(0,18){\circle{3}}
\end{picture}
&
\begin{picture}(10,30)(-5,-5) 
\put(2.1,0) {\circle{3}}
\put(2.1,1.8){\line(0,1){15}}
\put(2.1,18){\circle{3}}
\end{picture}
\end{array}\hspace{-2pt} {\Big \rangle}~,
$$
where $|\begin{picture}(25,6)(-3,-3)
\put(0,0) {\circle{3}}
\put(1.9,0.1){\line(1,0){15}}
\put(18,0){\circle{3}}
\end{picture}\rangle = 
|\hspace{-3pt}\uparrow\downarrow\rangle{-}
|\hspace{-3pt}\downarrow\uparrow\rangle~$.
Such plaquettes cover only one half of the possible elementary plaquettes 
of the lattice since two plaquettes cannot have a common side. 
In this way one necessarily has to break translation invariance and the 
resulting GS is fourfold degenerate in the thermodynamic limit,
in agreement with the Haldane's hedgehog argument described in 
Ref.~\cite{haldane}.

In the past, among several attempts to  guess the nature
of the non magnetic phase of this model, the description closest to ours
was that proposed by Zithomirski and Ueda \cite{zith}.
Amazingly, part of their conclusions were based on 
an unfortunate mistake in the series expansion \cite{columnar}.

The quantitative estimate of the order parameter can be obtained by
taking {\em first} the thermodynamic limit $N\to \infty$
of the order parameter 
$O_P(\delta)=\langle \hat{O}\rangle_{\delta}/N$
at fixed field $\delta$,  and then letting $\delta \to 0$,
$\lim\limits_{\delta \to 0} O_P(\delta)=O_P$ being the value of the
order parameter.
In order to estimate $O_P(\delta)$ at fixed size we have 
used the Hellmann-Feynmann theorem with a finite difference 
estimate  of $-de(\delta)/d\delta
\sim (e(0)-e(\delta))/\delta $. As shown in the inset of Fig.~\ref{chitra},
the finite size effects of this quantity seem to saturate
above the $N=64$ lattice size for $\delta \ge 0.04$, allowing a rather
convincing estimate of the dimer order parameter as $O_P \sim 0.1$, 
being sizably non zero.
The sharp crossover of the size effects for $N\ge 64$ 
is due to the presence of a finite triplet 
gap in the excitation spectrum (Fig.~\ref{gap}), implying, typically, 
a finite characteristic length.
The value of the order parameter $O_{P}$ is rather large considering
that $J_{2}/J_{1}=0.5$ is very close to the transition point
for the onset of sponaneous dimerization $J_{2}/J_{1}\simeq 0.40$. 
This is an interesting and measurable physical property that 
can be, in principle, investigated experimentally.

This work was partially supported by INFM (PRA HTCS and LOTUS) and 
MURST (COFIN99). 
We thank A. E. Trumper, M. Calandra, M. Capone, F. Becca,
G. Santoro, A. Parola and E. Tosatti, for fruitful discussions.

\end{document}